

\magnification = 1200
\overfullrule=0pt

\font\titlerm = cmr10 scaled\magstep 4
\font\titlerms = cmr7 scaled\magstep 4
\font\titlermss = cmr5 scaled\magstep 4
\font\titlei = cmmi10 scaled\magstep 4
\font\titleis = cmmi7 scaled\magstep 4
\font\titleiss = cmmi5 scaled\magstep 4
\font\titlesy = cmsy10 scaled\magstep 4
\font\titlesys = cmsy7 scaled\magstep 4
\font\titlesyss = cmsy5 scaled\magstep 4
\font\titleit = cmti10 scaled\magstep 4

\def\titlefont{\def\rm{\fam0\titlerm}
\def\it{\fam\itfam\titleit}
\textfont0 = \titlerm
\scriptfont0 = \titlerms
\scriptscriptfont0 = \titlermss
\textfont1 = \titlei
\scriptfont1 = \titleis
\scriptscriptfont1 = \titleiss
\textfont2 = \titlesy
\scriptfont2 = \titlesys
\scriptscriptfont2 = \titlesyss
\textfont\itfam = \titleit
\rm}

\def\sectionfont{\def\rm{\fam0\tenrm}
\def\it{\fam\itfam\tenit}
\def\bf{\fam\bffam\tenbf}
\textfont0 = \tenrm
\scriptfont0 = \sevenrm
\scriptscriptfont0 = \fiverm
\textfont1 = \teni
\scriptfont1 = \seveni  \scriptscriptfont1=\fivei
\textfont2 = \tensy
\scriptfont2 = \sevensy
\scriptscriptfont2 = \fivesy
\textfont\itfam = \tenit
\textfont\bffam = \tenbf
\rm}

\font\teenyfont = cmr5

\global\baselineskip = 1.2\baselineskip
\global\parskip = 4pt plus 0.3pt
\global\nulldelimiterspace = 0pt

\predisplaypenalty 1000


\def\endignore{}
\def\ignore #1\endignore{}

\newcount\dflag
\dflag = 0


\def\monthname{\ifcase\month
\or Jan \or Feb \or Mar \or Apr \or May \or June%
\or July \or Aug \or Sept \or Oct \or Nov \or Dec
\fi}

\def\datestamp{10 Jan 1994}




\def\endid{}
\def\id#1{\datestamp\hfill #1}

\def\endtitle{}
\def\title#1\endtitle{\vskip.15in\titlefont
\global\baselineskip = 2\baselineskip
#1\vskip.3in
\baselineskip = 0.5\baselineskip\sectionfont}

\def\lblfoot{This work was supported by the Director, Office of Energy
Research, Office of High Energy and Nuclear Physics, Division of High
Energy Physics of the U.S. Department of Energy under Contract
DE-AC03-76SF00098.}

\def\endauthors{}
\def\authors#1\endauthors{
#1\if\dflag = 0
\footnote{}{\noindent\lblfoot}\fi}

\def\endabstract{}
\def\abstract#1\endabstract{\vskip .2in%
\centerline{\sectionfont\bf Abstract}%
\vskip .1in%
\noindent#1%
\ifnum\dflag = 0
\footline = {\hfil}\pageno = 0
\vfill\eject
\pageno = 1\footline{\centerline{\sectionfont\folio}}
\fi\ifnum\dflag = 2
\footline = {\hfil}\pageno = 0
\vfill\eject
\fi}


\newcount\nsection
\newcount\nsubsection

\def\section#1{\global\advance\nsection by 1
\global\nsubsection = 0
\bigskip\noindent
\centerline{\sectionfont\bf\number\nsection.\ #1}
\nobreak\medskip\sectionfont\nobreak}

\def\subsection#1{\global\advance\nsubsection by 1
\bigskip\noindent
\centerline{\sectionfont \it \number\nsection.\number\nsubsection.\ #1}
\nobreak\smallskip\rm\nobreak}

\def\appendix#1#2{\bigskip\noindent%
\sectionfont \bf Appendix #1.\ #2
\nobreak\medskip\rm\nobreak}


\newcount\nref
\global\nref = 1

\def\ref#1#2{\xdef #1{[\number\nref]}
#1
\ifnum\nref = 1\global\xdef\therefs{\noindent[\number\nref] #2\ }
\else
\global\xdef\oldrefs{\therefs}
\global\xdef\therefs{\oldrefs\vskip.1in\noindent[\number\nref] #2\ }%
\fi%
\global\advance\nref by 1
}

\def\listrefs{\vfill\eject\section{References}\therefs}


\newcount\nfig
\global\nfig = 1

\def\fg#1\efig{\vskip .5in\noindent Fig.\ \number\nfig:\ #1%
\global\advance\nfig by 1}


\newcount\cflag
\newcount\nequation
\global\nequation = 1
\def\eqlabel{(1)}

\def\nexteqno{\ifnum\cflag = 0
\global\advance\nequation by 1
\fi
\global\cflag = 0
\xdef\eqlabel{(\number\nequation)}}

\def\lasteqno{\global\advance\nequation by -1
\xdef\eqlabel{(\number\nequation)}}

\def\label#1{\xdef #1{(\number\nequation)}
\ifnum\dflag = 1
{\escapechar = -1
\xdef\draftname{\teenyfont\string#1}}
\fi}

\def\clabel#1#2{\xdef\eqlabel{(\number\nequation #2)}
\global\cflag = 1
\xdef #1{\eqlabel}
\ifnum\dflag = 1
{\escapechar = -1
\xdef\draftname{\string#1}}
\fi}

\def\cclabel#1#2{\xdef\eqlabel{#2)}
\global\cflag = 1
\xdef #1{\eqlabel}
\ifnum\dflag = 1
{\escapechar = -1
\xdef\draftname{\string#1}}
\fi}


\def\eeq{}

\def\eqnn #1\eeq{$$ #1 $$}

\def\eq #1\eeq{\xdef\draftname{\ }
$$ #1
\eqno{\eqlabel \rlap{\ \draftname}} $$
\nexteqno}



\def\eol{& \eqlabel \rlap{\ \draftname} \crcr
\nexteqno
\xdef\draftname{\ }}

\def\eeol{& \eqlabel \rlap{\ \draftname}
\nexteqno
\xdef\draftname{\ }}



\def\eqa #1\eeq{\xdef\draftname{\ }
$$ \eqalignno{ #1 } $$
\global\cflag = 0}


\def\eg{{\it e.g.\/}}

\def\myinstitution{
    \centerline{\it Theoretical Physics Group}
    \centerline{\it Lawrence Berkeley Laboratory}
    \centerline{\it 1 Cyclotron Road}
    \centerline{\it Berkeley, California 94720}
}


\def\jref#1#2#3#4{{\it #1} {\bf #2}, #3 (#4)}

\def\NPB#1#2#3{\jref{Nucl.\ Phys.}{B#1}{#2}{#3}}
\def\PA#1#2#3{\jref{Physica}{#1A}{#2}{#3}}
\def\PLB#1#2#3{\jref{Phys.\ Lett.}{#1B}{#2}{#3}}

\def\PRD#1#2#3{\jref{Phys.\ Rev.}{D#1}{#2}{#3}}

\def\PRL#1#2#3{\jref{Phys.\ Rev.\ Lett.}{#1}{#2}{#3}}

\def\PRV#1#2#3{\jref{Phys.\ Rev.}{#1}{#2}{#3}}
\def\PTP#1#2#3{\jref{Prog.\ Theor.\ Phys.}{#1}{#2}{#3}}


\def\to{\mathop{\rightarrow}}


\def\frac#1#2{{{#1} \over {#2}}\,}  
\def\sfrac#1#2{{\textstyle\frac{#1}{#2}}}  


\def\Dsl{\hbox{/\kern-.6000em\rm D}} 



\def\scr#1{{\cal #1}}

\def\mybar#1{\kern 0.8pt\overline{\kern -0.8pt#1\kern -0.8pt}\kern 0.8pt}
\def\sla#1{\raise.15ex\hbox{$/$}\kern-.57em #1}
\def\Sla#1{\kern.15em\raise.15ex\hbox{$/$}\kern-.72em #1}

\def\roughly#1{\mathrel{\raise.3ex\hbox{$#1$\kern-.75em%
    \lower1ex\hbox{$\sim$}}}}


\def\tr{\mathop{\rm tr}}





\hyphenation{ba-ry-on ba-ry-ons ano-ma-ly ano-ma-lies}

\def\del{\delta}
\def\Del{\Delta}
\def\gam{\gamma}

\def\ep{\epsilon}

\def\Lam{\Lambda}

\def\Om{\Omega}
\def\sig{\sigma}
\def\Sig{\Sigma}

\def\ChPT{\raise.45ex\hbox{$\chi$}PT}

\def\hc{{\rm h.c.}}


\def\MeV{{\rm \ MeV}}
\def\GeV{{\rm \ GeV}}

\def\mkz{m_{K^0}}

\def\BB{\mybar B}
\def\chisim{$SU(3)_L \times SU(3)_R$}
\def\Lchi{\Lam_\chi}


\id
LBL-34779
\endid
\rightline{UCB-PTH-93/28}
\rightline{hep-ph/9401232}

\title
\centerline{Baryon Masses at Second Order}
\centerline{in Chiral Perturbation Theory}
\endtitle

\authors
\centerline{Richard F. Lebed {\it\ and\ } Markus A. Luty}
\footnote{}{\hskip-.26in\lblfoot}
\vskip .1in
\myinstitution
\endauthors

\abstract
We analyze the baryon mass differences up to second order in chiral
perturbation theory, including the effects of decuplet intermediate states.
We show that the Coleman--Glashow relation has computable corrections
of order $(m_d - m_u) m_s$.
These corrections are numerically small, and in agreement with the data.
We also show that corrections to the $\Sig$ equal-spacing rule are
dominated by electromagnetic contributions, and that the Gell-Mann--Okubo
formula has non-analytic corrections of order $m_s^2 \ln m_s$ which cannot
be computed from known matrix elements.
We also show that the baryon masses cannot be used to extract
model-independent information about the current quark masses.
\endabstract


\section{Introduction}

In this paper, we analyze the octet baryon masses at second order in chiral
perturbation theory.
Specifically, we expand isospin-violating mass differences to $O(e^2)$ and
$O((m_d - m_u) m_s)$, and strange mass differences to $O(m_s^2)$.
Our main motivation is to understand corrections to the famous $SU(3)$
relations for the baryon masses.

In chiral perturbation theory, the leading corrections to lowest-order
predictions are frequently non-analytic in the quark masses.
The non-analytic behavior in the chiral limit arises due to the
presence of massless Nambu--Goldstone bosons in intermediate states.
The non-analytic terms are therefore calculable if the necessary meson
couplings are known.
For example, the leading corrections to the Gell-Mann--Okubo formula are
$O(m_s^{3/2})$ and $O(m_s^2 \ln m_s)$
\ref\GMOcorr{L.-F. Li and H. Pagels, \PRL{26}{1204}{1971};
P. Langacker and H. Pagels, \PRD{10}{2904}{1974}.}.
However, we point out that the $O(m_s^2 \ln m_s)$ corrections depend on
two-derivative meson--baryon couplings which are not known accurately, so
only the $O(m_s^{3/2})$ corrections are calculable.
On the other hand, we show that the leading corrections to the
Coleman--Glashow relation are $O((m_d - m_u)m_s)$, and are calculable
in terms of well-measured quantities.
The predicted corrections agree with the data.
We also show that the $\Sig$ equal-spacing rule is dominated by
electromagnetic corrections, not by contributions from intermediate meson
states.

\section{Effective Lagrangian}

We will carry out our calculation in terms of an effective chiral lagrangian
\ref\effL{J. Schwinger, \PLB{24}{473}{1967};
S. Coleman, J. Wess, and B. Zumino, \PRV{177}{2239}{1969};
C. G. Callan, S. Coleman, J. Wess, and B. Zumino, \PRV{177}{2247}{1969};
S. Weinberg, \PA{96}{327}{1979}.}.
We will be brief in describing this lagrangian, since most of our notation
is standard.
The mesons are collected in the field
\eq
\xi(x) = e^{i\Pi(x) / f},
\eeq
which is taken to transform under $SU(3)_L \times SU(3)_R$ as
\eq
\xi \mapsto L \xi U^\dagger = U \xi R^\dagger.
\eeq
This equation implicitly defines $U$ as a function of $L$, $R$, and $\xi$.
\ignore
The meson fields are
\eq
\Pi = \frac 1{\sqrt 2}
\pmatrix{\frac 1{\sqrt 2}\pi^0 + \frac 1{\sqrt 6}\eta &
\pi^+ & K^+ \cr
\pi^- & -\frac 1{\sqrt 2} \pi^0 + \frac 1{\sqrt 6}\eta &
K^0 \cr
K^- & {\mybar K}^0 & -\frac 2{\sqrt 6} \eta \cr}.
\eeq
\endignore
The effective lagrangian is most conveniently written in terms of
\eq
V_\mu \equiv \frac i2\left(\xi \partial_\mu \xi^\dagger
+ \xi^\dagger \partial_\mu \xi\right), \qquad
A_\mu \equiv \frac i2\left(\xi \partial_\mu \xi^\dagger
-\xi^\dagger \partial_\mu \xi\right),
\eeq
which transform under \chisim\ as
\eq
V_\mu \mapsto U V_\mu U^\dagger + iU\partial_\mu U^\dagger, \qquad
A_\mu \mapsto U A_\mu U^\dagger.
\eeq
For any field $X$ transforming like $A_\mu$, the covariant derivative
\eq
\label\covder
\nabla_\mu X \equiv \partial_\mu X - i [V_\mu, X]
\eeq
transforms under \chisim\ as
\eq
\nabla_\mu X \mapsto U\nabla_\mu X U^\dagger.
\eeq

The chiral symmetry is broken explicitly by the quark masses and by
electromagnetism.
In the QCD lagrangian, the terms which explicitly break \chisim\ can be
written
\eq
\del\scr L = -(\mybar\psi_L m_q \psi_R + \hc)
+ e{\scr A}^\mu (\mybar\psi_L Q_L \gamma_\mu \psi_L
+ \mybar\psi_R Q_R \gamma_\mu \psi_R),
\eeq
where
\eq
\psi = \pmatrix{u \cr d \cr s \cr}, \qquad
m_q = \pmatrix{m_u &&\cr & m_d &\cr && m_s \cr}, \qquad
Q_L = Q_R = \pmatrix{-\frac 23 &&\cr & \frac 13 &\cr && \frac 13 \cr}.
\eeq
We note that the QCD lagrangian is formally invariant under
\chisim\ provided we assign to $m_q$, $Q_L$, and $Q_R$ the ``spurion''
transformation rules
\eq
m_q \mapsto L m_q R^\dagger, \quad
Q_L \mapsto L Q_L L^\dagger, \quad
Q_R \mapsto R Q_R R^\dagger.
\eeq
We therefore define the quantities
\eq
m_\pm \equiv \frac 12 (\xi^\dagger m_q \xi^\dagger \pm \hc), \qquad
Q_\pm \equiv \frac 12 (\xi^\dagger Q_L \xi \pm \xi Q_R \xi^\dagger).
\eeq
These quantities have definite parity and transform under \chisim\ as
\eq
m_\pm \mapsto U m_\pm U^\dagger, \qquad
Q_\pm \mapsto U Q_\pm U^\dagger.
\eeq

The simple transformation rules of the fields defined above make it easy
to write down the effective lagrangian.
For example, the leading terms involving only the mesons (which respect
$C$, $P$, and $T$) can be written
\eq
\scr L_\Pi = f^2 \tr(A^\mu A_\mu) + af^2\Lchi \tr(m_+)
+ h \frac{e^2 f^2 \Lchi^2}{16\pi^2} \tr(Q_+^2).
\eeq
The factors of $f = 93 \MeV$ and $\Lchi \sim 1 \GeV$ have been inserted
to ensure that the chiral expansion is an expansion in powers of
$m_q / \Lchi$ \ref\NDA{H. Georgi and A. V. Manohar, \NPB{234}{189}{1984}}.

Heavy fields such as baryons can be included in the effective lagrangian
as long as one only uses the lagrangian to describe processes in which
the momentum transfer to the heavy particle is small compared to the
scale $\Lchi$.
In this case, it is convenient to write an effective lagrangian in which
the baryon fields create nearly on-shell particles
\ref\heavy{H. Georgi, \PLB{240}{447}{1990};
T. Mannel, W. Roberts, and Z. Ryzak, \NPB{368}{204}{1992}.}
\ref\JM{E. Jenkins and A. V. Manohar, \PLB{255}{558}{1991}.}.
The basic idea is to write the baryon momentum as $p = M_B v + k$,
where $M_B$ is the common octet baryon mass in the $SU(3)$ limit, and $v$
is chosen so that all of the components of the residual momentum $k$ are
small compared to $\Lchi$ for the process of interest.
Hence, $k / \Lchi$ acts as an expansion parameter for these processes.
The effective lagrangian is then labelled by $v$, the approximately
conserved baryon velocity.
We will include both octet and decuplet baryon states, since the decuplet
states are not much heavier than the octet states, and thus intermediate
decuplet states are not suppressed
\ref\JMdec{E. Jenkins and A. V. Manohar, \PLB{259}{353}{1991}.
For a review, see E. Jenkins and A. V. Manohar,  {\it Proceedings of the
Workshop on Effective Field Theories of the Standard Model\/}, edited by
U.-G. Mei\ss ner (World Scientific, 1992).}.
This lagrangian is written in terms of spinor octet fields $B$ and
Rarita--Schwinger decuplet fields $T^\mu$ transforming under \chisim\ as
\eq
B \mapsto U B U^\dagger, \qquad
T^{jk\ell} \mapsto {U^j}_m {U^k}_n {U^\ell}_p T^{mnp}.
\eeq

The lowest-order terms in the effective lagrangian involving octet baryon
fields are
\eq
\label\Beff
\eqalign{
\scr L_{B1} &= \tr\left(\mybar B iv\cdot\nabla B\right)
+ D \tr\left( \mybar B s^\mu \{ A_\mu, B \} \right)
+ F \tr\left( \mybar B s^\mu [ A_\mu, B ] \right) \cr
&\qquad + b_1 \tr\left( \mybar B m_+ B \right)
+ b_2 \tr\left( \mybar B B m_+ \right) \cr
&\qquad + d_1 \frac{e^2 \Lchi}{16\pi^2} \tr(\BB Q_+^2 B)
+ d_2 \frac{e^2 \Lchi}{16\pi^2} \tr(\BB B Q_+^2) \cr
&\qquad + d_3 \frac{e^2 \Lchi}{16\pi^2} \tr(Q_+ \BB Q_+ B)
+ d_4 \frac{e^2 \Lchi}{16\pi^2} \tr(\BB Q_+) \tr(Q_+ B). \cr}
\eeq
(Terms involving $Q_-$ do not give rise to baryon masses at tree level,
and are ignored here.
We have omitted terms such as $\tr(m_+)\tr(\mybar BB)$ and
$\tr(Q_+^2)\tr(\mybar BB)$ which do not give rise to baryon mass
differences.
Terms containing an odd number of $Q_+$'s are forbidden by
charge conjugation invariance.)
The spin matrix $s^\mu$ is given by
$s^\mu \equiv (\gam^\mu - v^\mu \sla v) \gamma_5$,
and the covariant derivative acts on $B$ as in eq.\ \covder.

We now briefly consider the extraction of information about current
quark masses.
In chiral perturbation theory, we can only determine ratios of quark masses,
since the overall scale of the quark masses can be absorbed into unknown
coefficients in the chiral expansion.
Furthermore, baryon mass differences are unaffected by shifting the
current quark masses by a common constant.
Therefore, at leading order, the baryon mass differences are sensitive only
to the ratio of differences
\eq
\label\rdef
r \equiv \frac{m_d - m_u}{m_s - (m_u + m_d) / 2}.
\eeq
However, it is easy to see that the electromagnetic terms proportional to
$d_1$ and $d_2$ can exactly mimic the effect of a shift in $r$, and
therefore determination of $r$ is impossible without knowing $d_1$ and
$d_2$.
The computation of the electromagnetic contribution to the baryon mass
differences will be considered in a future publication
\ref\ourem{M. A. Luty and R. Sundrum, in preparation.}.

The leading-order lagrangian involving the decuplet fields is
\eq
\label\Teff
\scr L_{T1} = -\mybar T^\mu iv\cdot\nabla T_\mu +
\Del \mybar T^\mu T_\mu +
\scr C (\mybar T^\mu A_\mu B + \hc) + \cdots,
\eeq
where the $SU(3)$ indices in the last term are contracted as
$\ep_{jmn} T^{jk\ell} {A^m}_k {B^n}_\ell$.
Here, $\Del \simeq 300 \MeV$ is the octet--decuplet mass difference.
The omitted terms describe couplings among the decuplet fields which play
no role in our analysis.

In order to work consistently to order $m_q^2$, we will need the terms in
the second-order chiral lagrangian which contribute to the baryon mass
differences at tree level.
The terms
\eq
\label\kintoo
\frac{k_0}{\Lchi} \tr(m_+) \tr(\BB iv\cdot\nabla B)
+ \frac{k_1}{\Lchi} \tr(\BB m_+, iv\cdot\nabla B)
+ \frac{k_2}{\Lchi} \tr(\BB iv\cdot\nabla B m_+)
\eeq
contribute to wavefunction renormalization of the baryons.
We can eliminate these terms by making a field redefinition
\eq
B' \equiv B + \frac{k_0}{2\Lchi} \tr(m_+) B
+ \frac{k_1}{2\Lchi} m_+B
+ \frac{k_2}{2\Lchi} Bm_+.
\eeq
The effective lagrangian expressed in terms of $B'$ contains no terms of the
form of eq.\ \kintoo.
We will work with the fields $B'$ in what follows, dropping the primes for
notational convenience.
There are also terms
\eq
\label\sigmaterms
\frac{\sig_1}{\Lchi} \tr(m_+^2) \tr(\BB B)
+ \frac{\sig_2}{\Lchi} \tr(m_+) \tr(m_+)\tr(\BB B),
\eeq
which do not give rise to mass differences, and terms
\eq
\label\sigmatermstoo
\frac{\ell_1}{\Lchi} \tr(m_+) \tr(\BB m_+ B)
+ \frac{\ell_2}{\Lchi} \tr(m_+) \tr(\BB B m_+),
\eeq
which can be absorbed into the terms proportional to $b_1$ and $b_2$ in
eq.\ \Beff\ at the order to which we are working.
The only nontrivial terms in the second-order effective lagrangian are
\eq
\label\Befftoo
\eqalign{
\scr L_{B2} =& \frac{c_1}{\Lchi} \tr(\BB m_+^2 B)
+ \frac{c_2}{\Lchi} \tr(\BB B m_+^2) \cr
&+ \frac{c_3}{\Lchi} \tr(m_+ \BB m_+ B)
+ \frac{c_4}{\Lchi} \tr(\BB m_+) \tr(m_+ B). \cr}
\eeq

To count independent parameters, we must take into account the following
version of the Cayley--Hamilton theorem, which holds for any $3 \times 3$
traceless matrix $X$:
\eq
\tr(\BB \{X^2, B\}) + \tr(\BB X B X) - \tr(\BB X)\tr(XB)
- \frac 12 \tr(X^2)\tr(\BB B) = 0.
\eeq
For $X = Q_+$, this immediately shows that one of the electromagnetic terms
in eq.\ \Beff\ is redundant.
One can also use this relation to eliminate one of the quark mass terms
in eq.\ \Befftoo, since the trace part of the mass matrix can be absorbed
into redefinitions of the lowest-order couplings $b_1$ and $b_2$ at the
order we are working.
The 7 independent baryon mass differences are therefore determined by
the quark mass ratio $r$ defined in eq.\ \rdef\ and 8 effective couplings:
2 $O(m_q)$ couplings, 3 independent $O(e^2)$ terms, and 3 independent
$O(m_q^2)$ terms.
Including the loop corrections gives rise to calculable corrections to
the tree-level relations which depend on additional parameters, as
discussed below.
One therefore expects 1 prediction if we include only $O(m_q)$ and
$O(e^2)$ terms, and no predictions if we also include the $O(m_q^2)$ terms
and the loop corrections.
However, we will see that the actual situation is rather different.

\section{Baryon Mass Relations}

If we include $O(m_q)$ and $O(e^2)$ terms, we obtain several well-known
relations among the baryon masses.
They are the Gell-Mann--Okubo formula
\ref\GMOref{M. Gell-Mann, \PRV{125}{1067}{1962};
S. Okubo, \PTP{27}{949}{1962}.}
\eq
\label\GMO
\Del_{\rm GMO} \equiv \frac 34 \Lam + \frac 14 \Sig -
\frac 12 (N + \Xi) = O(m_{u,d}) + O(e^2),
\eeq
and the Coleman--Glashow relation
\ref\CGref{S. Coleman and S. L. Glashow, \PRL{6}{423}{1961}.}
\eq
\label\CG
\Del_{\rm CG} \equiv \Sig^+ - \Sig^- + n - p + \Xi^- - \Xi^0 = 0.
\eeq
(We use the particle names to denote the corresponding masses.)
In addition, there is one combination of baryon masses which gets
contributions only from the electromagnetic terms (the ``$\Sig$
equal-spacing rule'' \CGref)
\eq
\label\sigone
\Del_\Sig \equiv (\Sig^+ - \Sig^0) - (\Sig^0 - \Sig^-) = 0(e^2).
\eeq
In the language of $SU(3)$ group theory, the Gell-Mann--Okubo relation is
broken only by operators transforming under the $\Del I = 0$ piece of a
{\bf 27}, the Coleman--Glashow relation by the $\Del I = 1$ piece of a
{\bf 10}, and the $\Sig$ equal-spacing rule by the $\Del I = 2$ piece of
a {\bf 27}.
These relations hold at this order because the large representations
required to break these relations do not appear.

The experimental values are
\eq
\label\delexp
\Del_{\rm GMO} \simeq +6.5 \MeV, \quad
\Del_{\rm CG} = -0.3 \pm 0.6 \MeV, \quad
\Del_\Sig = 1.7 \pm 0.2 \MeV.
\eeq
It can be checked that the leading contributions to these quantities from
higher-order terms in the effective lagrangian are
\eq
\Del_{\rm GMO} = O(m_s^2), \quad
\Del_{\rm CG} = O(m_{u,d} m_s^2) + O(e^2 m_s), \quad
\Del_\Sig = O(e^2) + O(m_{u,d}^2),
\eeq
At our present level of understanding, these corrections are not calculable
and can only be estimated using power-counting arguments.

By contrast, the loop corrections to these relations are calculable in
terms of coefficients in the effective lagrangian which may be
experimentally measured.
The loop corrections are generally non-analytic in $m_q$, and when they are
larger than the counterterm contributions, we can make a prediction.
Loop corrections of the form of fig.\ 1a with a vertex coming from the
$b_1$ and $b_2$ terms in eq.\ \Beff\ give rise to symmetry-breaking terms
that transform as an {\bf 8}, and so do not affect the relations considered
above.
Loop corrections of the form of fig.\ 1b with vertices coming from the
$D$ and $F$ terms in eq.\ \Beff\ give rise to baryon mass corrections of
the form
\eq
\label\non
\del M_B \sim \frac{m_\Pi^3}{16\pi f^2} +
\frac{\Del_B m_\Pi^2}{16\pi^2 f^2}\ln\frac{m_\Pi^2}{\mu^2}
= O(m_q^{3/2}) + O(m_q^2 \ln m_q).
\eeq
Here $m_\Pi$ is a meson mass and $\Del_B$ is a baryon mass difference.
The coefficients $D$ and $F$ can be determined from semileptonic hyperon
decays.
The baryon masses also receive contributions from loops involving vertices
from the higher-order lagrangian such as
\eq
\label\unknown
\del\scr L_{B2} = \frac c{\Lchi} \tr(\mybar B A^\mu B A_\mu).
\eeq
Loop corrections of the form of fig.\ 1a with a vertex from terms such as
these can give rise to corrections to baryon masses of the form
\eq
\del M_B \sim \frac{m_\Pi^4}{16\pi^2 f^2 \Lchi}\ln{m_\Pi^2}
= O(m_q^2 \ln m_q).
\eeq
Since the coefficients of terms such as these are not measured, the
$O(m_q^2 \ln m_q)$ non-analytic corrections cannot be computed in general.
The dependence on $m_q$ of graphs involving decuplet intermediate states is
more complicated, since these contributions also depend on the
decuplet--octet mass difference $\Del$.
In the limit $\Del\to 0$, these contributions have the form of
eq.\ \non.

We now discuss corrections to the mass relations in detail.
The leading loop corrections to the Gell-Mann--Okubo relation are
$O(m_s^{3/2})$ and $O(m_s^2 \ln m_s)$.
These have been recently discussed in
ref.\ \ref\Jmass{E. Jenkins, \NPB{368}{190}{1992}.}.
One can check that there are $O(m_s^2 \ln m_s)$ corrections to
$\Del_{\rm GMO}$ from loop diagrams involving terms with unknown
coefficients such as eq.\ \unknown, so these contributions are not
computable;
in any case, they are not expected to be much larger than the $O(m_s^2)$
corrections from the counterterms.
In ref.\ \Jmass, it was found that the $O(m_s^{3/2})$ corrections give
$\Del_{\rm GMO} \simeq 15 \MeV$.
This is in satisfactory agreement with experiment, since we expect
$O(m_s^2)$ terms to give $\Del_{\rm GMO} \sim 10 \MeV$.

The Coleman--Glashow relation has no $O(m_q^2)$ or $O(e^2)$ corrections
from tree-level terms, so the leading corrections come from loop effects.
Loop contributions from two-derivative terms such as the one in
eq.\ \unknown\ do not contribute to $\Del_{\rm CG}$, because these they
transform as {\bf 8}'s and {\bf 27}'s.
We can therefore compute the leading corrections from the couplings in the
lowest-order lagrangian, eqs.\ \Beff\ and \Teff.
The contribution from octet intermediate states is
\eq
\Del^8_{\rm CG} = \frac {(K^+)^2 - (K^0)^2}{8\pi^2 f^2}
\Bigl[ D^2 (\Xi - N) + 3DF (\Lam - \Sig) \Bigr]
+ O(m_{u,d}^2).
\eeq
This expression is $O(m_s(m_d - m_u))$;
in particular, it is {\it analytic} in the quark masses.
This arises as follows:
The loop corrections to $\Del_{\rm CG}$ are $O(m_s(m_d - m_u)\ln m_s)$,
where the logarithm involves the renormalization scale $\mu$.
Because there are no $O(m_q^2)$ counterterms for $\Del_{\rm CG}$, changing
$\mu$ changes the result by $O(m_q^3)$.
We therefore chose $\mu = \mkz$, which corresponds to neglecting
$O(m_q^3 \ln m_s)$ contributions.
With this choice, the logarithms can be expanded in meson
mass differences, giving rise to an analytic result.

We now consider the contributions from decuplet intermediate states.
Because the octet--decuplet mass splitting $\Del \simeq 300 \MeV$ is
of the order of strange baryon mass differences, we will not treat
$\Del$ as a small parameter.
This makes the expressions for the decuplet contribution more complicated:
\vfill\eject
\eq
\label\deccg
\eqalign{
\Del^{10}_{\rm CG} = \frac {\scr C^2}{32\pi^2 f^2} \Biggl\{
& (n - p) \Bigl[ G_1(\Sig^* - N, K) + 4 G_1(\Del - N, \pi) \Bigr] \cr
& + (\Xi^- - \Xi^0) \Bigl[
G_1(\Sig^* - \Xi, K) + 2 G_1(\Om - \Xi, K) \cr
& \qquad\qquad\qquad\qquad\quad
+ G_1(\Xi^* - \Xi, \eta) + G_1(\Xi^* - \Xi, \pi) \Bigr] \cr
& + \sfrac 13 (\Sig^+ - \Sig^-) \Bigl[
8 G_1(\Del - \Sig, K) + 2 G_1(\Xi^* - \Sig, K) \cr
& \qquad\qquad\qquad\qquad\quad
+ 3 G_1(\Sig^* - \Sig, \eta) + 2 G_1(\Sig^* - \Sig, \pi) \Bigr] \cr
& + \sfrac 13 (\Sig^{*+} - \Sig^{*-}) \Bigl[
2 G_1(\Sig^* - N, K) + 2 G_1(\Sig^* - \Xi, K) \cr
& \qquad\qquad\qquad\qquad\quad
- 3 G_1(\Sig^* - \Sig, \eta) - G_1 (\Sig^* - \Sig, \pi) \Bigr] \cr
& + \sfrac{20}3 (\Xi^{*0} - \Xi^{*-} - \Sig^{*+} + \Sig^{*-}) \Bigl[
G_1(\Del - \Sig, K) - G_1(\Del - N, \pi) \Bigr] \cr
& + \sfrac 13 (\Xi^{*-} - \Xi^{*0}) \Bigl[
2 G_1(\Xi^* - \Sig, K) - 3 G_1(\Xi^* - \Xi, \eta) \cr
& \qquad\qquad\qquad\qquad\quad
+ G_1(\Xi^* - \Xi, \pi) \Bigr] \cr
& + \sfrac 13 \left( (K^0)^2 - (K^+)^2 \right) \Bigl[
4 G_2(\Del - \Sig, K) + G_2(\Sig^* - N, K) \cr
& \qquad\qquad\qquad\qquad\qquad\quad
- G_2(\Sig^* - \Xi, K) + 2 G_2(\Xi^* - \Sig, K) \cr
& \qquad\qquad\qquad\qquad\qquad\quad
- 6 G_2(\Om - \Xi, K) \Bigr] \cr
& + r \Bigl[G_3(\Sig^* - \Sig, \eta) - G_3(\Sig^* - \Sig, \pi) \cr
& \qquad\qquad\qquad
- G_3(\Xi^* - \Xi, \eta) + G_3(\Xi^* - \Xi, \pi) \Bigr] \Biggr\}.}
\eeq
In writing this result, we have used $SU(3)$ relations valid to
$O((m_d - m_u)m_s)$ for the decuplet masses to eliminate the dependence on
the poorly-measured $\Del$ isospin splittings
\ref\Lebed{R. F. Lebed, LBL-34704/UCB-PTH-93/27.}.
The terms proportional to $r$ arise from $\pi^0$--$\eta$ mixing, which is
proportional to $r$.
Here
\eqa
G_1(M, m) &\equiv 2(M^2 - m^2) F(m / M)
+ (2M^2 - m^2) \ln\frac{m^2}{\mu^2} \eol
G_2(M, m) &\equiv -\frac 1M (M^2 - m^2) F(m / M)
- M \ln\frac{m^2}{\mu^2}, \eol
G_3(M, m) &\equiv \frac 2{3M} (M^2 - m^2)^2 F(m / M)
+ M (\sfrac 23 M^2 - m^2) \ln\frac{m^2}{\mu^2}, \eeol
\eeq
where
\eq
F(x) \equiv \cases{\displaystyle
\frac 1{\sqrt{1 - x^2}}
\ln\frac{1 + \sqrt{1 - x^2}}{1 - \sqrt{1 - x^2}} & for $x < 1$, \cr
\vphantom{|}&\cr
\displaystyle
\frac 2{\sqrt{x^2 - 1}} \tan^{-1} \sqrt{x^2 - 1} & for $x \ge 1$. \cr}
\eeq
In these expressions, $\mu$ is the renormalization scale.
In the limit $M \gg m$, we have
\eqa
\label\bigone
G_1(M, m) &\to (2M^2 - m^2) \ln\frac{4M^2}{\mu^2} - \frac 12 m^2, \eol
G_2(M, m) &\to -M \ln\frac{4M^2}{\mu^2}, \eol
\label\biglast
G_3(M, m) &\to M(\sfrac 23 M^2 - m^2) \ln\frac{4M^2}{\mu^2}
- \frac 16 M m^2, \eeol
\eeq
up to terms that vanish as $M \to \infty$.
This shows that the decuplet contributions decouple in the limit where
the octet--decuplet splitting $\Del$ gets large, since the only terms
which do not vanish as $\Del\to\infty$ are analytic in the quark masses.
In this limit, the decuplet contributions can be absorbed into counterterms
in an effective lagrangian which does not contain decuplet fields, and the
$\mu$ dependence in eqs.\ \bigone--\biglast\ simply renormalizes the
couplings in this effective lagrangian.
In the opposite limit $M \ll m$, we have
\eqa
G_1(M, m) &\to -m^2 \ln\frac{m^2}{\mu^2}, \eol
G_2(M, m) &\to \frac{\pi m}2, \eol
G_3(M, m) &\to \frac{2\pi m^3}3, \eeol
\eeq
up to terms that vanish as $M \to 0$.
In this limit, $\Del^{10}_{\rm CG}$ has the same non-analytic dependence on
the quark masses as the contributions from octet intermediate states.
Changing the renormalization scale $\mu$ changes $\Del_{\rm CG}^{10}$ by
$O(m_q^3)$, so we again take $\mu = \mkz$ for purposes of numerical
evaluation.

Numerically, we find
\eq
\Del_{\rm CG}^{\rm theory} =
-2.2 D^2 + 1.3 DF + \scr C^2(0.5 \pm 0.5 + 8.1 r) \MeV.
\eeq
The numerical uncertainty in the coefficient of $\scr C^2$ is due to the
uncertainty in the decuplet isospin splittings.
We use the lowest-order fit values for $D$ and $F$, since for these values
the non-analytic corrections appear to be under control
\ref\Martin{M. A. Luty and M. White, LBL-34040/CfPA-TH-93-10, to be
published in {\it Phys.\ Lett.} {\bf B}.}:
$D = 0.85\pm 0.06$, $F = 0.52\pm 0.04$.
The errors are based on estimates of higher-order corrections, and may be
larger.
The coupling $\scr C$ can be determined from decuplet strong decays
to be $|\scr C| = 1.2 \pm 0.1$
\ref\nonlep{M. N. Butler, M. J. Savage, and R. P. Springer,
\NPB{399}{69}{1993}.}.
The value for $r$ is currently rather controversial, since it is closely
related to the problem of whether $m_u = 0$
\ref\KM{See \eg\ D. B. Kaplan and A. V. Manohar, \PRL{56}{2004}{1986};
H. Leutwyler, \NPB{337}{108}{1990}.}.
We will use the range $0.025 \leq r \leq 0.043$ which allows $m_u = 0$
(upper value) as well as the value from lowest-order chiral perturbation
theory (lower value).
We then find
\eq
\Del_{\rm CG}^{\rm theory} = 0.2 \pm 0.7 \MeV
\eeq
where the quoted error is dominated by the uncertainty on the decuplet
isospin splittings.
This prediction is in agreement with the experimental result
$\Del_{\rm CG} = -0.3 \pm 0.6 \MeV$.
We note that there is substantial cancellation between the octet and
decuplet contributions:
the octet contribution alone would give
$\Del_{\rm CG}^{\rm theory} = -1.0 \pm 0.2 \MeV$.

We also considered corrections to the $\Sig$ equal-spacing rule.
We find that all loop contributions to $\Del_\Sig$ are at most
$O(m_{u,d}^2)$, and are numerically negligible compared to the
experimental value.
This can be understood from the fact that the only $\Del I = 2$ operators
formed from the quark mass matrix have coefficient $(m_d - m_u)^2$.
We therefore conclude that $\Del_\Sig$ is dominated by the electromagnetic
contribution, which is expected to be of order
\eq
\del M_B^{\rm EM} \sim \frac{e^2 \Lchi}{16\pi^2} \sim 0.5 \MeV.
\eeq
This is clearly the right order-of-magnitude to explain the experimental
value.

\section{Conclusions}

We have considered the chiral expansion of the baryon mass differences to
$O(m_q^2)$ and $O(e^2)$ in the chiral expansion.
We have found that we cannot extract model-independent information about
the current quark masses, and that the $O(m_s^2 \ln m_s)$ corrections
to the Gell-Mann--Okubo relation are not calculable.
On the other hand, we showed that there are calculable corrections
to the Coleman--Glashow relation which are in agreement with the data,
and that the corrections to the $\Sig$ equal-spacing rule are dominated
by electromagnetic contributions.

\section{Acknowledgements}

We would like to thank R. Rattazzi, R. Sundrum, and M. Suzuki for useful
discussions.
This work was supported by the Director, Office of Energy Research, Office
of High Energy and Nuclear Physics, Division of High Energy Physics of the
U.S.\ Department of Energy under Contract DE-AC03-76SF00098.

\listrefs
\bye